\documentclass[romanappendices,journal,compsoc]{IEEEtran}

\usepackage{subfigure}
\usepackage{graphicx}
\usepackage{amsmath,amssymb,amsfonts}
\usepackage[breaklinks=true,colorlinks,bookmarks=false]{hyperref}
\usepackage{multirow}
\usepackage[table,xcdraw]{xcolor} 
\usepackage{ragged2e}

\graphicspath{{Figures/}}

\begin{document}
\title{Deep Learning and Computer Vision for Glaucoma Detection: A Review}

\author{Mona~Ashtari-Majlan, 
        Mohammad Mahdi~Dehshibi, David~Masip,~\IEEEmembership{Senior~Member,~IEEE}
        \IEEEcompsocitemizethanks{\IEEEcompsocthanksitem \protect M. A. and D. M. are with the Department of Computer Science, Multimedia, and Telecommunications, Universitat Oberta de Catalunya, Barcelona, Spain. M. M. D. is with the Department of Computer Science and Engineering, Universitat Carlos III de Madrid, Madrid, Spain. M.A. (e-mail: ashtari.mona@gmail.com; mashtarimajlan@uoc.edu).}
        \thanks{Manuscript \#TPAMI-XYZ received XYZ; revised XYZ.}
        }



\IEEEtitleabstractindextext{
\justify
\begin{abstract}

Glaucoma is the leading cause of irreversible blindness worldwide and poses significant diagnostic challenges due to its reliance on subjective evaluation. However, recent advances in computer vision and deep learning have demonstrated the potential for automated assessment. In this paper, we survey recent studies on AI-based glaucoma diagnosis using fundus, optical coherence tomography, and visual field images, with a particular emphasis on deep learning-based methods. We provide an updated taxonomy that organizes methods into architectural paradigms and includes links to available source code to enhance the reproducibility of the methods. Through rigorous benchmarking on widely-used public datasets, we reveal performance gaps in generalizability, uncertainty estimation, and multimodal integration. Additionally, our survey curates key datasets while highlighting limitations such as scale, labeling inconsistencies, and bias. We outline open research challenges and detail promising directions for future studies. This survey is expected to be useful for both AI researchers seeking to translate advances into practice and ophthalmologists aiming to improve clinical workflows and diagnosis using the latest AI outcomes.
\end{abstract}

}

\maketitle
\IEEEraisesectionheading{\section{Introduction}\label{sec:Introduction}}
\IEEEPARstart{G}{laucoma} is the second leading cause of irreversible blindness worldwide, affecting over 70 million people as of 2020~\cite{quigley2006number}. If left untreated, glaucoma leads to permanent vision loss due to damage to the optic nerve head and retinal nerve fiber layer~\cite{JONAS20172183}. However, despite improved understanding and management of glaucoma, it still accounts for approximately 10\% of global blindness~\cite{tham2014global,steinmetz2021causes}. This high disease burden motivates the development of enhanced diagnostic techniques to enable early diagnosis and timely intervention to prevent or slow down the further deterioration of vision.

Accurately diagnosing glaucoma remains challenging for several reasons~\cite{dervisevic2016challenges}. Firstly, glaucoma is often asymptomatic in its early stages, impeding detection without comprehensive eye examinations. Secondly, current diagnostic modalities like imaging tests and functional assessments have limitations in sensitivity and specificity. For example, while optical coherence tomography effectively captures structural changes in retinal layers, it cannot detect early functional loss. In contrast, visual field testing assesses functional impact but has low sensitivity for structural changes. Finally, the wide variability in glaucoma presentation, from subtle early symptoms to severe late-stage damage, makes definitive diagnoses difficult~\cite{JONAS20172183}. The complexity and subjectivity of evaluating diverse examination findings further complicate diagnosis.

This survey aims to provide a comprehensive overview of applying deep learning and computer vision algorithms to enhance glaucoma diagnosis\footnote{When we use the term ``diagnosis” in the context of papers that focus on deep learning and computer vision, we are referring to the use of AI technology to support medical diagnosis.} by overcoming these challenges. Such automated techniques offer the potential for earlier detection, more consistent quantification of progression, and ultimately preserving vision that would otherwise be lost to glaucoma~\cite{muhammad2017hybrid,hagiwara2018computer}. Synthesizing recent techniques, results, and open problems can deliver value to both ophthalmology practitioners and AI researchers. For ophthalmology practitioners, it highlights cutting-edge research to improve diagnostics accuracy and integrate intelligent systems into clinical workflows. For AI researchers, it provides a landscape analysis of the state-of-the-art, remaining gaps, and future opportunities to advance glaucoma diagnosis algorithms.

The remainder of the paper is organized as follows: Section~\ref{Sec:Search} outlines the employed search protocol to select relevant papers for review, ensuring comprehensive topic coverage. Section~\ref{Sec:Term} discusses clinical terminologies and definitions used in ophthalmology, enabling AI researchers to grasp the essential concepts necessary for interpreting the paper. Section~\ref{sec:Data} delves into the datasets used for training and testing deep learning and computer vision models and the performance metrics employed to evaluate their effectiveness. Section~\ref{sec:Feature} explores the various types of features, including structural, statistical, and hybrid, extracted for glaucoma diagnosis. Section~\ref{sec:Classification} reviews the latest research on developing end-to-end deep learning models for glaucoma diagnosis, categorizing them based on the type and architecture of the models. This section highlights the ability of deep learning models to integrate multiple modalities, analyze complex data, and provide accurate diagnosis and monitoring. Section~\ref{sec:Prediction} focuses on the methods proposed for early glaucoma prediction, shedding light on the advancements in prognostic techniques. In Section~\ref{sec:Challenges}, challenges and potential future directions are discussed, addressing the limitations and paving the way for further research in the field. Finally, Section~\ref{sec:Conclusion} serves as the conclusion, summarizing the key findings and contributions. It emphasizes the transformative potential of deep learning and computer vision in revolutionizing glaucoma diagnosis and management.

\section{Search protocol}\label{Sec:Search}
A systematic search was conducted to identify relevant studies on deep learning and computer vision techniques for glaucoma diagnosis published between 2017 and 2023. This date range was selected to capture the state-of-the-art advancements in this rapidly progressing field.

The following scholarly databases were searched: Web of Science, PubMed, IEEE Xplore, and Google Scholar. Targeted search terms included ``glaucoma," ``deep learning," ``computer vision," ``machine learning," ``artificial intelligence," and `` medical diagnosis" keywords. These were combined using Boolean operators and customized search strings tailored to each database. When available, searches were restricted to titles, abstracts and author keywords to filter potentially relevant papers efficiently. We also gave priority to leading conferences and journals in deep learning, computer vision, medical imaging, and ophthalmology. 


After deduplication, the records retrieved underwent two-phase screening. Title/abstract screening evaluated relevance to glaucoma diagnosis using deep learning and computer vision approaches. The full-text review then confirmed papers met the inclusion criteria: (1) written in English; (2) primary focus on automated glaucoma diagnosis/screening using deep learning or computer vision; (3) rigorous machine learning experiments and substantial technical depth. Studies were excluded if they: (1) focused on other ocular diseases, even with glaucoma sub-analysis; (2) primarily contributed clinically focused insights. We also excluded review papers, case reports, and conference abstracts given limited methodological detail.

Additionally, the reference lists of included papers were manually searched to identify any additional relevant studies that might have been missed in the initial search. This way, we could include seminal papers even if published outside the target venues. Data extraction from the selected papers involved capturing key information such as study design, dataset characteristics, deep learning architectures, evaluation metrics, and major findings. This information serves as the foundation for synthesizing the current state of research in glaucoma diagnosis using deep learning and computer vision.

While this search protocol aimed to be comprehensive, some relevant studies may have been inadvertently omitted, given the rapid pace of research in this field. Nonetheless, the systematic selection aimed to identify a representative sample for assessing the state-of-the-art deep learning and computer vision techniques for glaucoma diagnosis.

\section{Glaucoma: Definition and Diagnosis}\label{Sec:Term}

Familiarity with key ophthalmic concepts and terminologies~\cite{tielsch1991racial} is necessary, specifically for AI researchers, to effectively interpret glaucoma diagnosis research. This section defines relevant terms and tests used in clinical glaucoma assessment, with a summary provided in Table~\ref{tbl:terminologies}.

\begin{table*}[!htbp]
\centering
\caption{Ophthalmic terminologies.}
\label{tbl:terminologies}
\resizebox{\textwidth}{!}{%
\begin{tabular}
{m{3.5cm}m{2cm}m{9cm}}
\hline
\textbf{Terminology} & \textbf{Abbreviation} & \textbf{Definition}   \\ \hline

Retinal Ganglion Cells       & RGCs & Neurons in the innermost layer of the retina which receive and transmit visual information to the brain. \\

Retinal Nerve Fiber Layer    & RNFL & Layer of RGCs' nerve fibers (\textit{i.e.}, axons) that comprises the optic nerve and extends into the retina. \\

Ganglion Cell Layer & GCL &  Layer of RGCs' cell bodies. \\

Inner Plexiform Layer & IPL & Layer of RGCs' dendrites. \\

Ganglion Cell Complex & GCC & Forms the combination of RNFL, GCL, and IPL layers. \\

Ganglion Cell with the Inner Plexiform Layer & GCIPL & Layer that consists of GCL and IPL. \\

Optic Nerve Head & ONH & The structure in the posterior section of the eye that enables the exit of axons of RGCs and the entry/exit of blood vessels.\\

Neuroretinal Rim     & NRR & Area of the optic nerve head composed of retinal ganglion cell axons.\\

Optic Disc & OD & The optic disc and optic nerve head are interchangeable terms. \\

Optic Cup & OC & Central depression in the optic nerve head. \\

Cup-to-Disc Ratio & CDR  & Quantitative measure comparing the size of the optic cup to the optic disc. \\

Intraocular Pressure         & IOP  & Fluid pressure inside the eye, influenced by the balance of aqueous humor production and drainage.                 \\ \hline
\end{tabular}%
}
\end{table*}

\begin{figure}[!htpb]
    \centering
    \includegraphics[width=1\linewidth]{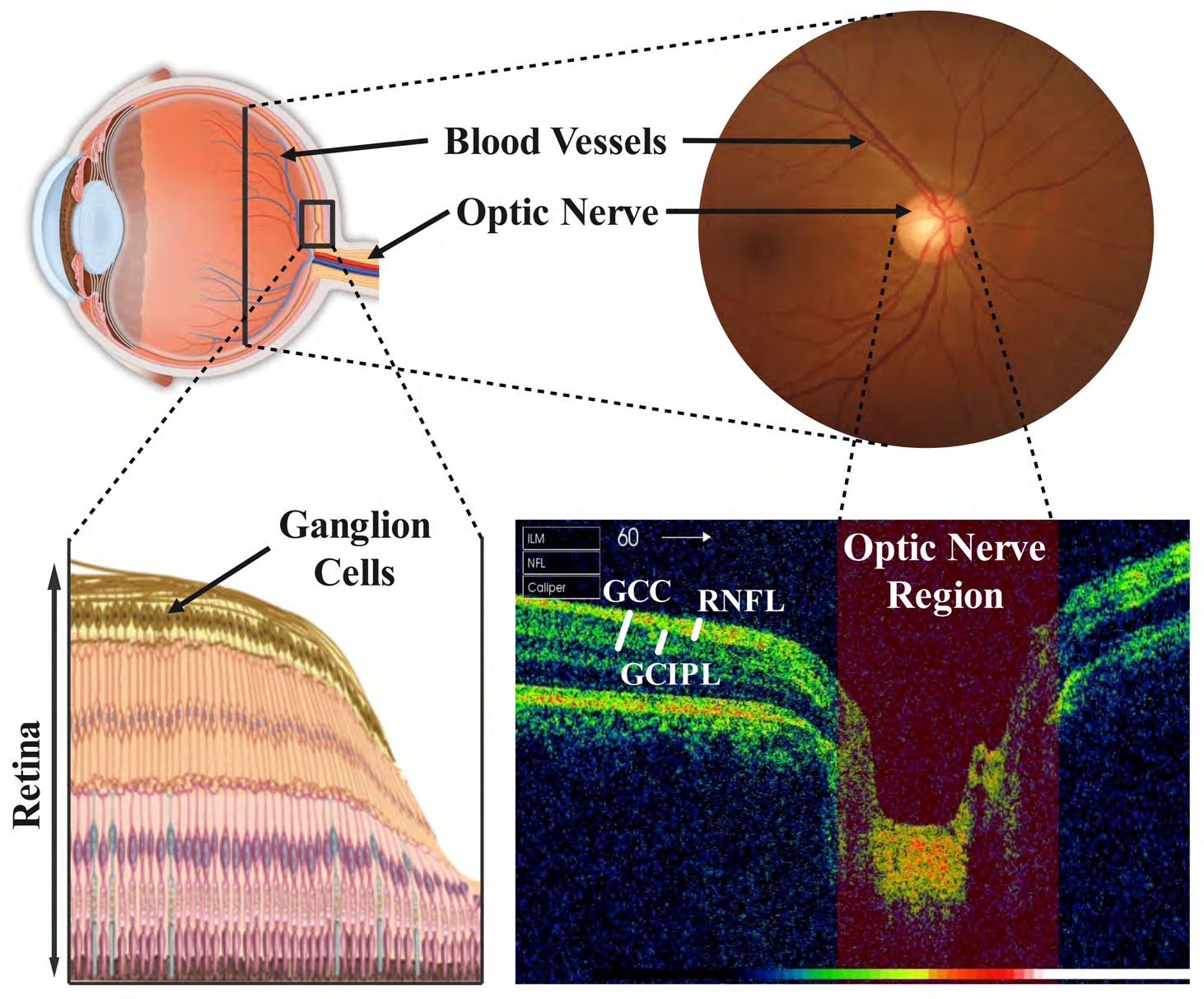}
    \caption{Anatomical structures of the human eye and optic nerve relevant to glaucoma detection. [Left] Schematic views, [Right] Fundus and OCT views~\cite{RAJA2020105342}. This figure was created using images licensed under Creative Commons.}
    \label{fig:OCT-RNFL}
\end{figure}

Glaucoma is a condition characterized by the degeneration of Retinal Ganglion Cells (RGCs), leading to structural changes in the retina~\cite{Weinreb2014Review}. These changes manifest as (1) the thinning of the Retinal Nerve Fiber Layer (RNFL), Ganglion Cell with the Inner Plexiform Layer (GCIPL), and Ganglion Cell Complex (GCC) profiles, (2) narrowing Neuroretinal Rim (NRR), and (3) cupping of the Optic Nerve Head (ONH) or enlargement of the Cup-to-Disc Ratio (CDR)~\cite{JONAS20172183} (see Fig.~\ref{fig:OCT-RNFL}). In addition to these structural changes, glaucoma also causes functional damage, resulting in defects in visual field sensitivity~\cite{ran2021deep}.

The process of detecting glaucoma is both complex and time-consuming~\cite{gutierrez2023artificial}. To gain valuable insights into the structural and functional changes associated with the disease, medical examinations and clinical expertise are utilized, where imaging techniques play a vital role. The National Institute for Health and Care Excellence in the UK recommends the use of fundus imaging to examine the ONH~\cite{coan2022automatic}.

Fundus imaging is a method that takes detailed images of the retina and OD. This helps with evaluating the appearance of the optic nerve, detecting vascular changes, and identifying any abnormalities. This imaging modality is a useful diagnostic tool for a variety of ocular conditions, including glaucoma. Optical Coherence Tomography (OCT) is another non-invasive imaging technique that produces highly detailed cross-sectional images of the retina, optic nerve, and other eye structures. It enables clinicians to visualize the retinal layers, measure the thickness of the RNFL, and assess the integrity of the NRR.

While fundus and OCT imaging techniques are commonly used to capture structural changes associated with glaucoma, the Humphrey Visual Field analysis (VF) measures the sensitivity of the visual field. This test, based on standard automated perimetry, allows patients to respond to visual stimuli presented at different locations within their visual field. By mapping the patient's responses, clinicians can identify visual field defects associated with glaucoma and assess the extent of retinal sensitivity loss to provide better ground truth~\cite{lee2019machine, TEKOUABOU2022115975, li2020development}.

The integration of fundus, OCT, and VF modalities~\cite{chen2019combination, Song2021Deep}, in addition to expert-level features~\cite{CHAI2018147} and biomarkers such as Intraocular Pressure (IOP)~\cite{ibrahim2022glaucoma, XUE2022104233, Dixit2021Assessing}, can leverage a broader range of inputs to improve the accuracy and enhance the performance of deep learning models, enabling more precise detection, prediction, and management of glaucoma.

\section{Datasets and Evaluation Metrics}\label{sec:Data}
In this section, we present a concise overview of the datasets used in the studies reviewed within this paper, as well as the evaluation metrics employed to assess the performance of trained models in glaucoma diagnosis. To facilitate ease of reference and enhance clarity, we have summarized this information in a table format (refer to Table~\ref{tab:datasets}\footnote{The links to the databases were active at the time of submitting this paper. In case of any inactive links, please contact the authors of the respective papers for updated information.}).

Various evaluation metrics have been used to assess the performance of glaucoma diagnostic models. To account for the lack of standardization in metrics across datasets, researchers have used both subject-level and finer-grained spatial metrics in their studies. Equation~\ref{eq:01} illustrates the commonly used quantitative measures in the reviewed studies. These include accuracy (ACC), sensitivity (SEN), specificity (SPE), precision (PRC), and the F1-score. Here, $TP$, $TN$, $FP$, and $FN$ represent true positive, true negative, false positive, and false negative, respectively. Additionally, the area under the receiver operating characteristic curve (AUC) metric has been employed to evaluate the performance of glaucoma diagnosis.

\begin{equation}
    \label{eq:01}
    \begin{split}
        \mathrm{Accuracy~(ACC)} = &~\frac{TP+TN}{TP+TN+FP+FN}. \\ 
        \mathrm{Sensitivity~(SEN)} = &~\frac{TP}{TP+FN}, \\
        \mathrm{Specificity~(SPE)} = &~\frac{TN}{TN+FP}, \\
        \mathrm{Precision~(PRC)} = &~\frac{TP}{TP+FP}, \\
        \mathrm{F1-score} = &~\frac{2\times \mathrm{SEN}\times \mathrm{SPE}}{\mathrm{SEN}+\mathrm{SPE}}.
    \end{split}
\end{equation}

\begin{table*}[!htb]
\caption{A review of the most commonly used datasets for glaucoma diagnosis. GT: Ground Truth, G: Glaucoma, H: Healthy. }
\label{tab:datasets}
\resizebox{\textwidth}{!}{%
\begin{tabular}
{m{2.25cm}m{1cm}m{1cm}m{1cm}m{2cm}m{3cm}m{3cm}m{1cm}}

\hline
\multirow{2}{*}{\textbf{Dataset}} & \multicolumn{3}{c}{\textbf{Number of Images}} & \multirow{2}{*}{\textbf{Resolution}} & \multirow{2}{3cm}{\textbf{Ground Truth Description}} & \multirow{2}{*}{\textbf{Note}} & \multirow{2}{*}{\textbf{Address}} \\ \cline{2-4}
& \textbf{Glaucoma} & \textbf{Healthy} & \textbf{Total} & & & \\ \hline

 &  &  &  &  &  &  &  \\
\multicolumn{8}{c}{\textbf{Fundus}} \\ \hline

 REFUGE~\cite{ORLANDO2020101570} & 121 & 1,079 & 1,200 & 2,124$\times$2,056, 1,634$\times$1,634 & Subject-level label, Segmentation GT (OC/OD) & -
 & \href{https://refuge.grand-challenge.org/}{Link}  \\
  
 LAG~\cite{Li2020Large} & 4,878 & 6,882 & 11,760 & 500$\times$500 & Subject-level label, Attention GT map & 5,824 out of all images (2,392 G/3,432 H) have Attention GT maps & \href{https://github.com/smilell/AG-CNN}{Link} \\
  
 RIM-ONE DL~\cite{ImageAnalStereol2346}
 & 172 & 313 & 485 & Various & Subject-level label & Combination of RIM-ONE-r1, r2 and r3~\cite{fumero2011rim}, Images are cropped at ON & \href{https://github.com/miag-ull/rim-one-dl}{Link} \\
  
 ORIGA~\cite{Zhang2010ORIGA} & 168 & 482 & 650 & Various & Subject-level label, Segmentation GT (OC/OD) & - &  Available upon request \\
  
 DRISHTI-GS1~\cite{sivaswamy2015comprehensive} & 70 & 31 & 101 & 2,896$\times$1,944 & Subject-level label, Segmentation GT (OC/OD), CDR Values, Notching & DRISHTI-GS1 is an extension of DRISHTI-GS~\cite{Sivaswamy2014Drishti} &  \href{http://cvit.iiit.ac.in/projects/mip/drishti-gs/mip-dataset2/Home.php}{Link} \\ 
 
 ACRIMA~\cite{diaz2019cnns}
 & 396 & 309 & 705 & 2,048$\times$1,536 & Subject-level label & Images are cropped at ON & \href{https://figshare.com/articles/dataset/CNNs_for_Automatic_Glaucoma_Assessment_using_Fundus_Images_An_Extensive_Validation/7613135}{Link} \\
  
 SIGF~\cite{Li2020DeepGF} & - & - & 3,671
 & Various & Subject-level label & 405 Sequential fundus images for glaucoma forecast including an average of 9 images per eye & \href{https://github.com/XiaofeiWang2018/DeepGF}{Link} \\

 HRF~\cite{budai2013robust} & 15 & 15 & - & 3,504$\times$2,336 & Subject-level label, Vessel GT & It also contains 15 images from diabetic retinopathy patients & \href{https://www5.cs.fau.de/research/data/fundus-images/}{Link} \\
 
 DRIONS-DB~\cite{CARMONA2008243}
 & - & - & 110 & 600$\times$400 & Contour of ON & Of all the images, 23.1\% belong to glaucoma patients and 76.9\% to eye hypertension patients & \href{http://www.ia.uned.es/~ejcarmona/DRIONS-DB/BD/DRIONS-DB.rar}{Link} \\ 
 
 ODIR-5K~\cite{odir5k} & 307 & 1,620 & - & Various & Subject-level label & Contains 5,000 images divided into eight categories
 & \href{https://www.kaggle.com/datasets/andrewmvd/ocular-disease-recognition-odir5k}{Link} \\
 
 JSIEC~\cite{cen2021automatic} & 13 & 54 &  & Various & Subject-level label & Contains 1087 images divided into 37 categories & \href{https://www.kaggle.com/datasets/linchundan/fundusimage1000}{Link} \\
 
 RIGA~\cite{Almazroa2017Optic} & - & - & 750 & Various & Segmentation GT (OC/OD) & No subject-level label & \href{https://deepblue.lib.umich.edu/data/concern/data_sets/3b591905z?locale=en}{Link} \\ 

 \hline

&  &  &  &  &  &  &  \\
\multicolumn{8}{c}{\textbf{OCT}} \\ \hline

AGE~\cite{FU2020101798} & - & - & 300 & Various & Subject-level label (Angle-closure Vs. Open-angle glaucoma)$\color{blue}^{\ddagger}$,
Scleral Spur localization & Each OCT volume contains 16 2D images resulting in 4,800 images (1:4 ratio of closure to open-angle samples) & \href{https://age.grand-challenge.org/}{Link} \\ 
 
\hline

&  &  &  &  &  &  &  \\
\multicolumn{8}{c}{\textbf{OCT \& Fundus}} \\ \hline

AFIO~\cite{RAJA2020105342} & 32 & 18 & 50 & OCT:  951$\times$456, Fundus: 2,032$\times$1,934 & Subject-level label, CDR values & The data is from 26 subjects & \href{https://data.mendeley.com/datasets/2rnnz5nz74/2}{Link} \\

\hline

&  &  &  &  &  &  &  \\
\multicolumn{8}{c}{\textbf{VF}} \\ \hline

Rotterdam~\cite{Kucur2018deep} & 2,279 & 244 & 2,523 & - & Subject-level label & 24-2 VF test pattern & \href{https://github.com/serifeseda/early-glaucoma-identification}{Link} \\

Budapest~\cite{Kucur2018deep} & 532 & 1,735 & 2,267 & - & Subject-level label & 24-2 VF test pattern & \href{https://github.com/serifeseda/early-glaucoma-identification}{Link} \\
 \hline
 
\multicolumn{8}{m{17.25cm}}{$\color{blue}^{\ddagger}$ Glaucoma can be classified into two broad categories, angle-closure and open-angle. The former is considered a more aggressive form of the disease compared to the latter~\cite{FU2020101798}.} \\ 

\end{tabular}%
}
\end{table*}

\section{Feature Extraction}\label{sec:Feature}

The feature extraction process involves deriving a set of representative features from the input data. These features can be extracted using various methods, such as conventional statistical or structural techniques (also known as hand-crafted features), deep learning architectures that automatically learn relevant features from the data, or biomarkers identified by domain experts based on their knowledge. The resulting informative features are then used as inputs to train the learning model.

\subsection{Structural Features}
Structural image measurements based on physical characteristics of the optic nerve head help clinicians in quantifying retinal structures relevant to glaucoma. These measurements provide objective information that can be used to determine the severity of glaucoma and monitor its progression. 
Clinical knowledge can help improve the accuracy and interpretability of machine learning algorithms\cite{xu2021hierarchical}. One of the most important structural measurements that can be extracted from fundus images and is used by many clinicians is the CDR.
A higher CDR indicates a larger Cup and a higher risk of glaucoma~\cite{Jiang2020JointRCNN}. Therefore, many glaucoma screening works focus on accurate OC and OD segmentation tasks~\cite{WANG2021107810, Fu2018Joint, juneja2020automated, Tabassum2020CDED, LIU2019103485, P2021107512, haleem2018novel, latif2022odgnet, bajwa2019two}. 

Soorya et al.~\cite{Soorya2018automated}, for instance, proposed an adaptive threshold framework for segmenting the OC and OD based on geometrical features. Following a clinical approach used by ophthalmologists, the proposed algorithm tracks blood vessels inside the Disc region, identifies the points at which different vessels are bent for the first time, and connects them to obtain the contours of the OC.  
They further calculated the vertical CDR (the distance between the topmost and bottommost points of the OC and OD) and proposed a threshold based on which they classified an image as normal/suspect glaucoma/glaucoma.
Mvoulana et al.~\cite{MVOULANA2019101643} proposed using the K-means clustering algorithm with an intensity-based nearness criterion for pixel classification, followed by a model-based boundary fitting approach employing the circular Hough transform for OC and OD segmentation. They later calculated the CDR value and used a threshold to classify healthy and glaucoma patients. The method achieved 98\% accuracy on the DRISHTI-GS1 dataset for the final glaucoma diagnosis.

Deep learning models have shown promising results in various image analysis tasks, including image segmentation~\cite{Minaee2022Image}. For example, Jiang et al.~\cite{Jiang2018Optic} proposed to use two Faster R-CNNs~\cite{NIPS2015_14bfa6bb} to segment the OC and OD separately and find the minimal bounding boxes for the two regions. The authors extended their work in~\cite{Jiang2020JointRCNN}, where they focused on the joint OC and OD segmentation problem by proposing JointRCNN, an end-to-end region-based convolutional neural network. The JointRCNN consists of four major parts: feature extraction module, Disc proposal network, Disc attention module and Cup proposal network.
The feature extraction module is shared by OC and OD segmentation tasks, and atrous convolution~\cite{chen2018encoder} is used to improve the feature extraction performance in this module.
The Disc and Cup proposal networks generate bounding box proposals, and the Disc attention module is proposed to connect the two networks. 
To improve the segmentation performance, Fu et al.~\cite{Fu2018Joint} proposed using polar transformation along with a multi-label deep network (M-Net) for joint segmentation of the OC and OD in retinal images. The proposed M-Net consists of a multi-scale input layer to construct an image pyramid, a U-shape convolutional network to learn the rich hierarchical representation, a side-output layer as an early classifier that produces a companion local prediction map for different scale layers, and a multi-label loss function to generate the final segmentation map. They calculated the CDR value for glaucoma screening and evaluated the performance of the proposed method on the ORIGA and Singapore Chinese Eye Study (SCES) datasets achieving an AUC of 85.08\% and 89.98\%, respectively.
Liu et al.~\cite{LIU2019103485} proposed a joint OC and OD segmentation method based on a semi-supervised model to take advantage of both labeled and unlabeled data to improve the segmentation performance.
The proposed conditional Generative Adversarial Nets (cGAN)-based architecture consists of a segmentation net, a generator and a discriminator to learn a mapping between the fundus images and the corresponding segmentation maps. Both the segmentation net and the generator in the proposed framework focus on learning the conditional distributions between fundus images and their corresponding segmentation maps. At the same time, the discriminator determines whether the image-label pairs come from the empirical joint distribution. Furthermore, the proposed method performed better than its fully-supervised version on both ORIGA and REFUGE datasets, with AUC values of 86.22\% and 90.11\% and accuracy rates of 76.57\% and 82.78\%, respectively.

While segmentation-based methods have shown effective performance, accurate CDR measurement remains challenging due to factors such as overlapping regions, low contrast between regions, shape variability and inhomogeneity of the OD, insufficient labels, and errors in intermediate steps. Zhao et al.~\cite{Zhao2020Direct} proposed a direct CDR estimation method based on a semi-supervised learning scheme that bypasses the intermediate segmentation step. The proposed two-stage cascaded approach consisted of two phases: unsupervised feature representation of fundus image with a Convolutional Neural Network (CNN) and CDR value regression using a Random Forest (RF) regressor. The proposed model achieved an AUC of 90.50\% for glaucoma diagnosis on a dataset of 421 fundus images.
Zhou et al.~\cite{zhou2019adaptive} proposed an adaptive weighted locality-constrained sparse coding approach for glaucoma diagnosis, which combines locality constraint with sparse constraint and employs a weighted locality constraint constructed by adaptively combining multiple distance measurements.
The proposed approach achieved an accuracy of 88.63\% and 85.56\% in diagnosing CDR for DRISHTI-GS and RIM-ONE-r2, respectively.

To predict the average RNFL thickness previously extracted from Spectral Domain OCT (SD-OCT\footnote{SD-OCT is a type of OCT that uses a faster scanning speed and higher resolution to produce more detailed images of the eye's internal structures compared to traditional time-domain OCT.}) scans and allow for quantification of neural damage, Medeiros et al.~\cite{MEDEIROS2019513} proposed a deep learning model based on fundus OD images. They used a private dataset containing 32,820 pairs of OD fundus images and SD-OCT scans to predict average RNFL thickness. They also assessed the ability of predicted and actual RNFL thickness values to discriminate between glaucomatous and healthy eyes and achieved an AUC of 94.00\% and 94.40\%, respectively.
Raja et al.~\cite{Raja2021Clinically} proposed a novel approach to objectively grade glaucoma as an early suspect or advanced stage based on the degeneration of RGCs. They first segmented the RNFL, GCIPL, and GCC regions and extracted their thickness information.
Subsequently, they employed a Support Vector Machine (SVM) to evaluate the severity of glaucoma, achieving a high accuracy of 91.17\% on the AFIO dataset.
Lee et al.~\cite{lee2020diagnosing} also proposed a deep learning framework based on NASNet~\cite{Zoph_2018_CVPR} for diagnosing glaucoma using SD-OCT images. They extracted features from the GCIPL thickness map, GCIPL deviation map, RNFL thickness map, and RNFL deviation map and fed the extracted features into the deep learning classifier for glaucoma diagnosis. 
The proposed model achieved an AUC of 99.00\% with a sensitivity of 94.7\% and a specificity of 100.0\% on a private dataset with 350 glaucomatous and 307 healthy SD-OCT image sets.

\subsection{Statistical Features}

Statistical features such as intensity-based, texture-based, morphological, and wavelet-based features are used to extract quantitative measurements from the medical data~\cite{semmlow2021biosignal}. In glaucoma diagnosis, these features can help differentiate between normal and abnormal eyes.
Claro et al.~\cite{CLARO2019102597} conducted an extensive study to determine the best set of features for fundus image representation, which included Local Binary Pattern, Gray Level Co-occurrence Matrix (GLCM)~\cite{haralick1973textural}, Histogram of Oriented Gradients (HOG)~\cite{dalal2005Histograms}, Tamura~\cite{tamura1978textural}, Gray Level Run Length Matrix (GLRM)~\cite{GALLOWAY1975172}, morphology, and seven CNN architectures, yielding a 30,682-D feature vector. They then used the gain ratio algorithm for feature selection and concluded that a combination of the GLCM and pre-trained CNNs has the best glaucoma diagnosis accuracy of 92.78\% on 1675 images of DRISHTI-GS, RIM-ONE, HRF, JSIEC, and ACRIMA datasets.

Juneja et al.~\cite{JUNEJA2022117202} extracted GLRM and GLCM features from the wavelet-filtered OCT images, along with 18 other statistical features. Thereafter, discriminative features were selected using the gain ratio, information gain and correlation statistical methods. In addition, they used a 3D-CNN architecture to extract features and perform classification. They finally used majority voting and weighted decision fusion strategies to provide the final classification results taking into account K-nearest neighbour (k-NN), RF, SVM and the probability given by the 3D-CNN model. Based on the experimental results, the proposed framework achieved a precision of 95.00\%, sensitivity of 97.00\% and F1-score of 96\% on a private dataset of 1,110 OCT scans (847 glaucoma cases and 263 normal cases).
Maheshwari et al.~\cite{Maheshwari2017Automated} proposed using empirical wavelet transform (EWT) to decompose fundus images and obtaining correntropy features from decomposed EWT components for glaucoma diagnosis. 
These extracted features are then ranked based on the t-value feature selection algorithm and fed to a least-squares SVM for normal and glaucoma image classification, achieving an accuracy of 98.33\% and a specificity of 96.67\% on a private and a public database.
Nayak et al.~\cite{NAYAK2021102559} proposed an automatic feature extraction method based on a meta-heuristic optimization algorithm called a real-coded genetic algorithm. To extract high-level features from the fundus images directly, the proposed method adopts a strategy based on maximizing the inter-class distance and minimizing intra-class variability. The final feature vectors are then used in conjunction with an SVM classifier for glaucoma diagnosis. The experimental results on a dataset of 1,426 fundus images (589 normal and 837 glaucoma) yielded an accuracy of 97.20\%.

Extracting the Region of Interest (ROI) speeds up subsequent processing by excluding irrelevant image regions. As previously stated, the OD is an important ROI region in glaucoma diagnosis.
For example, Vin{\'i}cius dos Santos Ferreira et al.~\cite{VINICIUSDOSSANTOSFERREIRA2018250} proposed a framework for OD semantic segmentation based on the U-Net model. Texture features were then extracted from both the RGB channels and gray levels of the segmented region using phylogenetic diversity indexes. The proposed approach resulted in an accuracy of 98.50\%, with a sensitivity of 98.00\%, specificity of 100\%, F1-score of 96.00\%, and AUC of 98.10\% on the RIM-ONE, DRIONS-DB, and DRISHTI-GS datasets for glaucoma diagnosis.
Bisneto et al.~\cite{BISNETO2020106165} also proposed a cGAN with a U-Net generator and a PatchGAN discriminator for OD segmentation. They used cGAN in conjunction with taxonomic indexes to extract textural attributes from the segmented OD region for glaucoma classification. Three different classifiers, namely Multilayer Perceptron (MLP), Sequential Minimal Optimization, and RF, were utilized in diagnosing glaucoma on the RIM-ONE and DRISHTI-GS datasets. All of the classifiers achieved 100\% accuracy and AUC.

\subsection{Hybrid Features} 

Combining structural and statistical features can improve the performance of glaucoma diagnosis models by incorporating different perspectives. While structural features provide information about clinical measurements and anatomical changes in the retina, statistical features provide image-based information. As a result, more comprehensive information for effective glaucoma screening can be obtained by utilizing hybrid features. 
To this end, Balasubramanian and N.P.~\cite{BALASUBRAMANIAN2022109432} extracted structural and statistical features from fundus images and presented correlation-based feature selection algorithms as well as a Kernel-Extreme Learning Machine classifier for glaucoma diagnosis. To extract structural features, they first segmented OD and OC using a Fuzzy C-Means Clustering algorithm and calculated CDR and Cup shape features. They further extracted features like GLCM, Anisotropic Dual-Tree Complex Wavelet Transform (ADT-CWT)~\cite{SWAMIDOSS201314}, Fractal Texture Analysis~\cite{Cheung2009Quantitative}, SURF~\cite{bay2006surf}, Pyramid HOG~\cite{bosch2007representing}, Mean Gray-Level, Color Intensity Features, and Super-pixels. The experimental results demonstrated a maximum overall accuracy of 99.61\% with 99.89\% sensitivity and 100\% specificity on the public and private retinal fundus datasets containing 7,280 images.
Guo et al.~\cite{guo2020automated} proposed the increasing field of view (IFOV) feature model to fully extract textural, statistical, and other hidden image-based features. In the IFOV model, there are four different image scales, ranging from small to large: OC region, OD region, ROI (cropped image around OD), and global fundus image. 
They then extracted CDR and statistical features from images in different scales using the Gabor transform and GLCM, followed by feature selection using the adaptive synthetic sampling approach~\cite{He2008ADASYN}. Finally, the extracted features are used to train a gradient-boosting decision tree classifier for glaucoma screening resulting in 84.30\% and 83.70\% accuracy on ORIGA and DRISHTI-GS1 datasets, respectively.

Thakur and Juneja~\cite{THAKUR2020102137} presented a set of reduced hybrid features derived from structural and nonstructural features to classify the retinal fundus images. The structural features extracted included CDR and Disc damage likelihood scale. Whereas, non-structural features included GLRM, GLCM, First order statistical, Higher order spectra, Higher order cumulant and Wavelets. They performed feature selection using the wrapper approach and used different classifiers, including k-NN, Neural network (NN), RF, SVM, and Na\"ive Bayes (NB), for glaucoma diagnosis. Among all the classifiers, SVM exhibited the highest performance with an accuracy of 97.20\%, a specificity of 96.00\%, a precision of 97.00\%, and a sensitivity of 97.00\% on the DRISHTI-GS and RIM-ONE datasets.
Kausu et al.~\cite{KAUSU2018329} proposed a novel method for glaucoma identification based on time-invariant feature CDR and ADT-CWT features. They first segmented the OD using the Fuzzy C-Means clustering method, and the OC using Otsu's thresholding. An MLP model is finally used for glaucoma classification achieving an accuracy of 97.67\% with 98\% sensitivity. The dataset used in this paper was collected from the Venu Eye Institute \& Research Centre in New Delhi, India, and contained a total of 86 images, 51 of which were healthy and 35 from glaucoma patients.

The Inferior Superior Nasal Temporal (ISNT) rule is one of the widely used techniques for assessing structural damage to the optic nerve head in clinical practice. 
According to the ISNT rule, the thickness of the NRR in normal eyes decreases in the following order: inferior region $>$ superior region $>$ nasal region $>$ temporal region, whereas the NRR in glaucomatous optic discs violates this rule~\cite{Chan2013Diagnostic}.
Pathan et al.~\cite{PATHAN2021102244} extracted clinical features from the segmented OD and OC regions, including CDR estimation and ISNT rule verification in the NRR area. They further extracted color and texture features from the NRR area to analyze glaucoma-related changes in fundus images. Three color spaces were used to extract color features, including RGB, CIEL*a*b, and HSV, while textural features were extracted using the GLCM approach. 
Finally, to diagnose glaucoma from normal samples, they used an SVM, a three-layered NN, and an AdaBoost classifier with dynamic ensemble selection. The SVM algorithm demonstrated the highest accuracy of 95.00\% and 90.00\% on the DRISHTI-GS1 and a private dataset, respectively.
Singh et al.~\cite{singh2021enhanced} also extracted ISNT regions and CDR, along with 20 statistical features such as Homogeneity, Contrast, and Correlation. They used the combination of four machine learning algorithms (i.e., SVM, K-NN, NB, and a 3-layered MLP) and achieved 95.82\% sensitivity, 98.59\% specificity with an accuracy of 98.60\% on the DRIONS-DB dataset.
Martins et al.~\cite{MARTINS2020105341} segmented the OC and OD and calculated morphological features. The extracted features included CDR, vertical length CDR, Rim-to-disc area ratio, which also provides an interpretation of the ONH shape, and ISNT values and rule compliance. The proposed pipeline also included a glaucoma confidence level assessment using a classification network with a MobileNetV2~\cite{sandler2018mobilenetv2} feature extractor as a backbone.
The final decision combines the glaucoma confidence level with the calculated morphological features, yielding an accuracy of 87.00\%, sensitivity of 85.00\%, specificity of 88.00\%, and AUC of 93.00\% on a merged dataset of several publicly available datasets.

Different types of ophthalmic images provide information on retinal pathology from various angles, and combining them can aid in glaucoma diagnosis. Chen et al.~\cite{chen2019combination} proposed an automatic method for early glaucoma screening using Enhanced Depth Imaging OCT (EDI-OCT) and fundus images. The method includes segmenting the anterior lamina cribrosa surface in EDI-OCT images with a region-aware strategy and residual U-Net and extracting structural features such as lamina cribrosa depth and deformation. Similarly, in fundus images, scanning lines and brightness compensation are used to segment the OC and OD regions, and the CDR and textural features are extracted. Hybrid features that combine structural parameters from EDI-OCT and textural features from fundus images are then used for training and classification to screen glaucoma in the early stage using gcForest. The proposed method achieved an accuracy of 96.88\% with 91.67\% sensitivity on a private dataset.

\section{End-to-end Glaucoma Classification}
\label{sec:Classification}

End-to-end deep learning models have shown promising results in glaucoma classification, outperforming feature-based methods~\cite{Thompson2020Assessment, RAGHAVENDRA201841, HERVELLA2022108347, raghavendra2019two, ZHAO2022102295, Li_2019_CVPR, guo2022dsln, Dixit2021Assessing, xu2021hierarchical}. Because of their ability in incorporating holistic contextual information in the training process, end-to-end deep learning models can potentially reduce the risk of information loss and improve generalizability. For example, Hemelings et al.~\cite{hemelings2021deep} showed that end-to-end deep learning models take advantage of contextual information outside the optic nerve head region in fundus images to detect glaucoma and estimate the CDR. In this section, we review studies that mainly focus on these models for diagnosing glaucoma disease. We classify these models mainly based on their architectures into Convolutional Neural Networks, Autoencoder-based Networks, Attention-based Networks, Generative Adversarial Networks, Geometric Deep Learning Networks, and Hybrid Networks.

\subsection{Convolutional Neural Networks}

Convolutional neural networks (CNNs) have been widely used for glaucoma diagnosis~\cite{DEPERLIOGLU2022152, LI20181199, JUN2021115211, Zhao_Liao_Zou_Chen_Li_2019, RAN2019e172, Kucur2018deep}. In general, these models are used to extract higher-level features from raw image data, with earlier layers focusing on simple features such as colors and edges and later layers identifying more complex shapes and structures.
The CNN architecture consists of different types of layers, including convolutional, pooling, and fully connected layers.
Convolutional layers apply learned filters to the input image, generating activation feature maps that represent detected features of different complexity and level of detail. The pooling layer is used to reduce the dimensionality of the feature maps to decrease computational complexity. Fully connected layers finally map the extracted high-level features to the desired output classes. Many studies reviewed in this paper employed state-of-the-art CNN architectures, including Inception-v3~\cite{szegedy2016rethinking}, ResNet~\cite{He_2016_CVPR}, EfficientNet~\cite{Tan2019Efficient}, and DenseNet~\cite{Huang_2017_CVPR} pre-trained on ImageNet~\cite{russakovsky2015imagenet}. These studies either used the pre-trained architectures without further modifications or slightly modified them to fit the research's objectives. Additionally, some studies developed new CNN architectures entirely from scratch, tailoring the model design specifically for the glaucoma classification task. Table~\ref{tab:CNN-Arc} summarize the papers that used CNN for glaucoma classification.

\begin{table*}[!htbp]
    \caption{Comparison of papers utilizing CNN architectures for glaucoma classification. Studies with (*) mark tested their model on an auxiliary dataset. N: Normal, G: Glaucoma, S: Glaucoma Suspect, A: Advanced glaucoma, E: Early glaucoma, Avg: Average.}
    \label{tab:CNN-Arc}
    \resizebox{\textwidth}{!}{%
    \begin{tabular}[!htb]{m{1cm}m{2cm}m{1.5cm}m{3cm}m{1cm}m{1cm}m{1cm}m{1cm}m{1.1cm}m{1cm}m{0.75cm}}

    \hline
    \multirow{2}{*}{\textbf{Study}} & \multirow{2}{*}{\textbf{Model}} & \multirow{2}{*}{\textbf{Data type}} & \multirow{2}{*}{\textbf{Dataset}} & \multicolumn{6}{c}{\textbf{Classification Performance Measures (\%)}} & \multirow{2}{*}{\textbf{Code}} \\ \cline{5-10}
    &  & &  & \textbf{ACC} & \textbf{SEN} & \textbf{SPE} & \textbf{PRC} & \textbf{F1-score} & \textbf{AUC} &  \\ \hline
        
    \cite{RAGHAVENDRA201841} &  18-layer CNN & Fundus & Private (589 N, 837 G) & 98.13 & 98.00 & 98.30 & - & - & - & - \\ 
        
    \multirow{2}{1cm}{\cite{Wu2020Leveraging}*} & \multirow{2}{2cm}{EfficientNet-based} & \multirow{2}{2cm}{Fundus} & Private (1,586 G, 2,244 N), RIGA & 93.29 & 96.03 & 91.42 & - & - & 98.29 & \multirow{2}{2cm}{\href{https://github.com/WuJunde/Semi-supervised-Glaucoma-Detection}{Link}} \\ 
    & & & LAG & 95.81 & 98.40 & 94.22 & - & - & 99.49 & \\
        
    \cite{LI20181199} & Inception-v3 & Fundus & Private (29,466 N, 2,620 S, 7,659 G)
    & - & 95.60 & 92.00 & - & - & 98.60 & - \\ 

    \cite{JUN2021115211} & TRk-CNN & Fundus & Private (403 N, 208 S, 381 G) & 88.94 (Avg) & 85.37 (S Vs. N), 90.36 (G Vs. N) & 89.33 (Avg) & 74.47 (S Vs. N), 94.94 (G Vs. N) & 79.55 (S Vs. N), 92.59 (G Vs. N) & - & - \\
        
    \cite{shibata2018development} & ResNet-18 & Fundus & Private (Training: 1,424 G, 1,818 N) & - & - & - & - & - & 96.50 & - \\
        
    \cite{Ahn2018deep} & 3-layer CNN & Fundus & Private (1,542, 786 N, 467 A, 289 E) & 87.90 & - & - & - & - & 94.00 & -\\
        
    \cite{Zhao_Liao_Zou_Chen_Li_2019} & WSMTL & Fundus & ORIGA & - & - & - & - & - & 92.00 & - \\
        
    \cite{Liao2020Clinical} & EAMNet & Fundus & ORIGA & - & - & - & - & - & 88.00 & - \\
        
    \cite{XUE2022104233} & ResNet & Fundus, VF & Private (1,695 N, 1,201 mild, 1,607 moderate, 1,868 severe) & 85.50 (G Vs. N), 82.50 (Multi-class$\color{blue}^{\ddagger}$)
    & - & - & - & - & 88.10 (G Vs. N), 95.60 (Multi-class) & - \\
        
    \multirow{2}{2cm}{\cite{Kucur2018deep}*} & \multirow{2}{2cm}{7-layer CNN} & \multirow{2}{2cm}{VF} & Rotterdam & - & - & - & 87.40, 98.60 & - & - & \multirow{2}{2cm}{\href{https://github.com/serifeseda/early-glaucoma-identification}{Link}} \\ 

    &  &  & Budapest & - & - & - & 98.60 & - & - &  \\ 
        
    \multirow{2}{2cm}{\cite{WANG2020101695}*} & \multirow{2}{2cm}{CNN} & \multirow{2}{2cm}{OCT} & HK (2,926 G, 1,951 N) & 92.70 & - & - & - & 94.10 & 97.70 & \multirow{2}{2cm}{-} \\
    & & & Stanford (806 G, 425 N) & 86.00 & - & - & - & 88.90 & 93.30 & \\
        
    \cite{RAN2019e172} & ResNet-34 & OCT & Private (2,926 G, 1,951 N) & 91.00 & 89.00 & 96.00 & - & - & 96.90 & - \\ 
        
    \cite{Thompson2020Assessment} & ResNet-34 & OCT & Private (612 G, 542 N) & - & - & - & - & - & 96.00 & - \\ 

    \multirow{2}{2cm}{\cite{Thakoor2021Robust}*} & \multirow{2}{2cm}{Inception-v3} & \multirow{2}{2cm}{OCT} & Private (192 G, 545 N) & 90.40 & - & - & - & - & - & \multirow{2}{2cm}{\href{https://github.com/LIINC/TCAV4OCT}{Link}} \\

    &  &  & Private (57 G, 44 N) & 91.10 & - & - & - & - & - & \\      
        
    \cite{ASAOKA2019136} & 6-layer CNN & OCT & Private (1,579 G, 359 N)
    & - & - & - & - & - & 93.70 & - \\ 
        
    \hline
    
    \multicolumn{11}{m{18.5cm}}{$\color{blue}^{\ddagger}$ Severity classification between mild, moderate and severe glaucoma} \\ 

    \end{tabular}%
    }
\end{table*}

Wang et al.~\cite{WANG2020101695} proposed an end-to-end semi-supervised multi-task learning CNN for classifying OCT B-scan images as glaucoma or normal and investigating the relationship between structural and functional changes in glaucoma eyes. The proposed CNN comprises three components: a shared feature extraction module with a ResNet-18 backbone, a glaucoma classification module, and a VF measurement regression module. To develop and test the proposed method, they also created one of the largest glaucoma OCT image datasets (\textit{i.e.}, HK dataset) with 975,400 B-scans from 4,877 volumes.
Zhao et al.~\cite{Zhao_Liao_Zou_Chen_Li_2019} also proposed a Weakly-Supervised Multi-Task Learning (WSMTL) method for accurate evidence identification, OD segmentation, and automated glaucoma diagnosis. WSMTL consists of a skip and densely connected CNN for multi-scale feature representation of fundus structure, a pyramid integration structure for generating high-resolution evidence maps, a constrained clustering branch for OD segmentation, and a fully-connected discriminator for automated glaucoma diagnosis. 
The model is trained on weakly labeled data with binary diagnostic labels (normal/glaucoma), and the output is a pixel-level segmentation mask and glaucoma diagnosis label.
Liao et al.~\cite{Liao2020Clinical} proposed a clinically interpretable CNN architecture (EAMNet) for glaucoma diagnosis. EAMNet aggregates the features extracted from a CNN backbone with ResBlock at various scales to bridge the gap between semantic and localization information at multiple scales in order to improve glaucoma diagnosis accuracy.
Additionally, EAMNet generates refined evidence activation maps that highlight the glaucoma-specific discriminative regions recognized by the network, aiming to provide a more transparent interpretation.

Xue et al.~\cite{XUE2022104233} proposed a three-phased framework for 1) screening, 2) detecting glaucoma from normal, and 3) classifying glaucoma into four severity levels. 
In the first phase, they used IOP to screen out patients with glaucoma.
Two distinct ResNet architectures, namely DetectionNet and ClassificationNet, were trained independently during the second and third phases, respectively. In the case of DetectionNet, the fusion of fundus and Voronoi VF images served as input for the classification of normal and glaucoma cases. Conversely, ClassificationNet utilized a Voronoi VF image as input to categorize glaucoma severity into mild, moderate, or severe.
Jun et al.~\cite{JUN2021115211} also proposed a Transferable Ranking Convolutional Neural Network (TRk-CNN) for the multi-class classification of normal, glaucoma suspect, and glaucoma eyes.
TRk-CNN employs DenseNet as the backbone CNN model and combines the weights of the primitive classification model to reflect inter-class information to the final classification phase, where there is a high correlation between classes.

\subsection{Autoencoder-based Networks}

An autoencoder is a type of neural network that uses an encoder-decoder structure to extract important features from input data. The encoder maps the input into a lower-dimensional latent space, and the decoder maps the latent representation back to the original input space.
This mechanism allows the autoencoder to capture meaningful data representations, which has significant potential for improving glaucoma detection methodologies. For instance, Raghavendra et al.~\cite{raghavendra2019two} proposed a two-layer sparse autoencoder to extract effective and important features from fundus images for glaucoma detection. The proposed network consists of two cascaded autoencoders for unsupervised feature learning and a Softmax layer for supervised glaucoma classification. Table~\ref{tab:AE-Arc} summarizes the papers that used an autoencoder-based architecture for glaucoma diagnosis.

\begin{table*}[!htbp]
    \caption{Comparison of papers utilizing autoencoder-based architectures for glaucoma classification. Studies with (*) mark tested their model on an auxiliary dataset. SAE: Sparse Autoencoder, N: Normal, G: Glaucoma.}
    \label{tab:AE-Arc}
    \resizebox{\textwidth}{!}{%
    \begin{tabular}[!htb]{m{1cm}m{2cm}m{1.5cm}m{3cm}m{1cm}m{1cm}m{1cm}m{1cm}m{1.1cm}m{1cm}m{0.75cm}}
        \hline
        \multirow{2}{*}{\textbf{Study}} & \multirow{2}{*}{\textbf{Model}} & \multirow{2}{*}{\textbf{Data type}} & \multirow{2}{*}{\textbf{Dataset}} & \multicolumn{6}{c}{\textbf{Classification Performance Measures (\%)}} & \multirow{2}{*}{\textbf{Code}} \\ \cline{5-10}
         &  & &  & \textbf{ACC} & \textbf{SEN} & \textbf{SPE} & \textbf{PRC} & \textbf{F1-score} & \textbf{AUC} &  \\ \hline
        
        \multirow{2}{2cm}{\cite{HERVELLA2022108347}*} & \multirow{2}{2cm}{U-Net-based} & \multirow{2}{2cm}{Fundus} & REFUGE & -  & - & - & - & - & 97.60 & - \\
        & & & DRISHTI-Gs & - & - & - & - & - & 94.74 & - \\
        
        \cite{pascal2022multi} & U-Net-based & Fundus & REFUGE & - & - & - & - & - & 96.76 & - \\
        
        \cite{raghavendra2019two} & SAE & Fundus & Private (589 N, 837 G) & 95.30 & 95.20 & - & 96.80 & 95.00 & - & - \\
        
        \cite{Pal2018Eyenet} & G-EyeNet & Fundus & DRIONS-DB & - & - & - & - & - & 92.30 & - \\
        
        \cite{Raja2021Clinically} & RAG-Net\textsubscript{v2} & OCT & AFIO & 94.91 & 97.14 & 91.66 & 94.44 & 95.77 & 98.71 & \href{https://github.com/taimurhassan/rag-net-v2}{Link} \\
        \hline
    \end{tabular}%
    }
\end{table*}

Pal et al.~\cite{Pal2018Eyenet} proposed a deep learning model (G-EyeNet) for detecting glaucoma using an encoder-decoder framework. G-EyeNet comprises an encoder, decoder, and classifier module. The encoder-decoder structure is used for image reconstruction and unsupervised feature learning, while the classifier module uses the latent-space distribution learnt by the encoder to classify glaucoma. The framework is trained using multi-task learning to minimize reconstruction and classification losses.
Raja et al.~\cite{Raja2021Clinically} proposed a hybrid autoencoder-based convolutional network framework (RAG-Net\textsubscript{v2}) for segmenting RGC regions and classifying glaucoma.
RAG-Net\textsubscript{v2} is first trained to extract RGC regions, particularly the RNFL, GCIPL, and GCC regions. 
The RAG-Net\textsubscript{v2} encoder end is then used to perform RGC-aware classification of healthy and glaucoma samples.

Hervella et al.~\cite{HERVELLA2022108347} proposed a multi-task approach for simultaneous OD and OC segmentation and glaucoma classification. The proposed network consists of an encoder-decoder structure that is shared between tasks and takes advantage of both pixel-level and image-level labels during network training.
In addition, they used a multi-adaptive optimization strategy to ensure that both tasks contribute equally to parameter updates during training, avoiding the use of loss weighting hyperparameters.
Pascal et al.~\cite{pascal2022multi} proposed a multi-task deep learning model to detect glaucoma in retinal fundus images while segmenting the OD and OC and locating the fovea.
The proposed model is trained using a U-Net encoder-decoder convolutional network as a backbone architecture and is adapted to handle the four tasks using independent optimizers. The multi-task model outperforms the single task of detecting glaucoma because it leverages related tasks and their similarities to achieve better performance.
Ren et al.~\cite{ren2020task} also proposed a task decomposition framework based on an encoder-decoder architecture for both semantic segmentation of OC and OD and glaucoma classification. The three subsequent subtasks that they proposed are (1) pixel-wise semantic segmentation of fundus images, (2) prediction of OD and OC instance class labels, and (3) classification of glaucoma and normal fundus images. The framework used a sync-regularization to penalize the deviation between the outputs of pixel-wise semantic segmentation and the instance class prediction tasks and outperformed the single-task model.

\subsection{Attention-based Networks} \label{subsec:attention}

The attention mechanism in deep learning enables models to selectively focus on the most important parts of input data, enhancing prediction accuracy and efficiency. Using attention in glaucoma screening models intuitively satisfies the need to focus on the key pathological areas rather than other redundant information. 
In the reviewed literature, the term ``attention" is used to describe two distinct concepts. The first concept is attention in terms of models focusing on specific parts of input data through the use of auxiliary information, such as attention maps obtained from domain experts~\cite{Li_2019_CVPR, Li2020Large}, heatmaps generated by Grad-CAM~\cite{George2020IEEE}, and segmenting disease-related regions in the input~\cite{Jiang2020JointRCNN, Raja2021Clinically, HERVELLA2022108347}.
The second concept refers to the attention-based architectures, which are also the basis of Vision Transformers (ViT)~\cite{Dosovitskiy2021Transformers}. We further discuss these concepts in the following, and a summary of the reviewed paper is presented in  Table~\ref{tab:ATT-Arc}.

Li et al.~\cite{Li_2019_CVPR} proposed an attention-based CNN for glaucoma detection (AG-CNN), which highlights salient regions by incorporating attention maps to remove redundancy from fundus images. 
The proposed model comprises three subnets: an attention prediction subnet, a pathological area localization subnet and a glaucoma classification subnet. The attention maps in this paper were collected in the large-scale attention-based glaucoma (LAG) database through a simulated eye-tracking experiment. The ophthalmologists provided attention maps by focusing on the salient regions of the fundus images. The LAG database includes 5,824 fundus images labeled with either positive glaucoma or negative glaucoma. They extended their work in~\cite{Li2020Large} to enlarge the LAG database to 11,760 fundus images. In addition to the supervised method presented in~\cite{Li_2019_CVPR}, this paper further proposed a weakly supervised learning to incorporate attention maps in a weakly supervised manner for glaucoma detection.
George et al.~\cite{George2020IEEE} proposed an end-to-end attention-guided 3D CNN model for glaucoma detection and visual field index (VFI) estimation using high-resolution 3D OCT volumes. The proposed model consists of three pathways sharing the same network architecture. The first pathway takes the raw 3D-OCT cube as input and learns global retinal structures relevant to glaucoma detection and VFI estimation. Similarly, the inputs of the other two pathways are computed during training, guided by the 3D Grad-CAM~\cite{Selvaraju2017GradCAM} attention heatmaps. Each pathway outputs the class label, and the entire model is trained concurrently to minimize the sum of the three losses. The final output is obtained by fusing the predictions of the three pathways.

\begin{table*}[!htbp]
    \caption{Comparison of papers utilizing attention-based architectures for glaucoma classification. Studies with (*) mark tested their model on an auxiliary dataset. N: Normal, G: Glaucoma, S: Glaucoma Suspect, A: Advanced glaucoma, E: Early glaucoma, Avg: Average, Att: Attention module.}
    \label{tab:ATT-Arc}
    \resizebox{\textwidth}{!}{%
    \begin{tabular}[!htb]{m{1cm}m{2cm}m{1.5cm}m{2cm}m{1cm}m{1cm}m{1cm}m{1cm}m{1.1cm}m{1cm}}
        \hline
        \multirow{2}{*}{\textbf{Study}} & \multirow{2}{*}{\textbf{Model}} & \multirow{2}{*}{\textbf{Data type}} & \multirow{2}{*}{\textbf{Dataset}} & \multicolumn{6}{c}{\textbf{Classification Performance Measures (\%)}} \\ \cline{5-10}
         &  & &  & \textbf{ACC} & \textbf{SEN} & \textbf{SPE} & \textbf{PRC} & \textbf{F1-score} & \textbf{AUC} \\ \hline

        \multirow{3}{2cm}{\cite{ZHAO2022102295}*} & \multirow{3}{2cm}{CNN + Att} & \multirow{3}{2cm}{Fundus} & LAG & 97.12 & 95.20 & 98.16 & - & 95.47 & 99.28 \\
        & & & REFUGE & 95.25 & 80.00 & 96.94 & - & 78.82 & 93.32 \\
        & & & RIM-ONE & 93.96 & 89.74 & 97.12 & - & 90.91 & 97.19 \\
        
        \cite{zhao2020egdcl} & CNN + Att & Fundus & LAG & 97.12 & 97.21 & 97.07 & - & 96.65 & 99.31 \\
        
        \cite{GUO2022106910} & CNN + Att & Fundus & LAG & 96.70 & 96.10 & 97.00 & - & 95.00 & 99.60 \\

        \multirow{7}{2cm}{\cite{WasselICPR2022}} & Swin~\cite{Liu2021Swin} & \multirow{7}{2cm}{Fundus} & \multirow{7}{2cm}{6 publicly available datasets$\color{blue}^{\ddagger}$}
        & 93.20 & 92.57 & 93.43 & - & - & 97.77 \\
        & Cait~\cite{Touvron_2021_ICCV} & & & 94.50 & 89.92 & 96.04 & - & - & 97.90 \\
        & CrossViT~\cite{Chen2021CrossViT} & & & 94.30 & 86.73 & 96.94 & - & - & 96.87 \\
        & XciT~\cite{NEURIPS2021_a655fbe4} & & & 93.55 & 88.6 & 95.23 & - & - & 97.2 \\
        & ResMlp~\cite{Touvron2023ResMLP} & & & 91.50 & 85.94 & 93.43 & - & - & 96.00 \\
        & DeiT~\cite{touvron2021training} & & & 88.00 & 81.70 & 90.10 & - & - & 94.60 \\
        & ViT~\cite{Dosovitskiy2021Transformers} & & & 87.40 & 77.20 & 90.60 & - & - & 92.60 \\
        & BeiT~\cite{bao2021beit} & & & 85.50 & 83.82 & 86.15 & - & - & 92.70 \\
        
        \multirow{2}{2cm}{\cite{xu2021automatic}*} & \multirow{2}{2cm}{TIA-Net} & \multirow{2}{2cm}{Fundus} & Private (1,005 N, 877 G) & 85.70 & 84.90 & 86.90 & - & - & 92.90 \\
        & & & ORIGA & 76.60 & 75.30 & 77.20 & - & - & 83.50  \\
        
        \multirow{2}{2cm}{\cite{Li_2019_CVPR}*} & \multirow{2}{2cm}{AG-CNN} & \multirow{2}{2cm}{Fundus} & LAG & 
        95.30 & 95.40 & 95.20 & - & 95.10 & 97.50 \\ 
        & & & RIM-ONE & 85.20 & 84.80 & 85.50 & - & 83.70 & 91.60 \\
        
        \cite{Li2020Large} & AG-CNN & Fundus & LAG & 92.20 & 95.40 & 96.70 & - & 95.40 & 98.30 \\
        
        \cite{garcia2021circumpapillary} & CNN + Att & OCT & Private (90 N, 72 E, 57 A) & 87.88 (Avg$\color{blue}^{\ddagger\ddagger}$)
        & 81.82 (Avg) & 90.91 (Avg) & 81.82 (Avg) & 81.82 (Avg) & -  \\
        
        \cite{Song2021Deep} & DRT & OCT, VF & Private (697 G, 698 N) & 88.30 & 93.70 & 82.40 & - & 88.90 & 93.90 \\ 
        
        \cite{George2020IEEE} & CNN + Grad-CAM & OCT & Private (427 N, 3,355 G) & 91.07 & 95.12 & - & 94.73 & 94.88 & 93.77 \\
        
        \hline
        
    \multicolumn{10}{m{16.25cm}}{$\color{blue}^{\ddagger}$ LAG, ODIR-5K, ORIGA, REFUGE, DRISHTI-GS1, HRF} \\ 
    \multicolumn{10}{m{16.25cm}}{$\color{blue}^{\ddagger\ddagger}$ Average categorical results for discriminating between healthy, early and advanced glaucoma samples.} \\  
    
    \end{tabular}%
    }
\end{table*}

Garc{\'\i}a et al.~\cite{garcia2021circumpapillary} proposed a hybrid neural network with hand-driven features and a deep learning backbone that has skip-connections to include tailored residual and attention modules to refine the automatic features of the latent space. They specifically fed the backbone model with raw OCT B-scans and used a descriptor to extract RNFL thickness-based information as hand-driven features. The model was trained using a few-shot learning technique for discriminating between healthy, early and advanced glaucoma scans.
Zhao et al.~\cite{ZHAO2022102295} proposed a student-teacher framework where both the student and teacher models have identical network architecture consisting of two separate attention pathways (\textit{i.e.,} spatial and channel attention modules). The spatial attention module identifies salient image regions and prune feature responses, while the channel attention module identifies salient feature channels to preserve the activations relevant to the glaucoma diagnosis task. The proposed framework aims to improve glaucoma diagnosis on imbalanced data by augmenting feature distribution with feature distilling and re-weighting.
Guo et al.~\cite{GUO2022106910} proposed a multitask teacher-student framework for unbiased glaucoma screenings and visualizations of model decision-making areas. The teacher network in the proposed framework utilizes a ResNet-34 backbone to extract semantic feature maps of different depths for constructing a multi-scale discrimination module and adopts the self-attention mechanism module to make the network pay attention to spatial information and channel information at the same time. This ensures the quality of the generated evidence map and provides reliable preliminary results for glaucoma classification. Meanwhile, the student network, incorporating a dual-branch CNN structure and a collaborative learning module, simultaneously performs glaucoma diagnosis and generates the corresponding evidence map.

The ViT is a deep learning architecture that utilizes a self-attention mechanism to capture the global characteristics of an image. ViT applies the transformer model, which was originally designed for natural language processing, to computer vision tasks~\cite{Dosovitskiy2021Transformers}. Wassel et al.~\cite{WasselICPR2022} evaluated the performance of more than seven different ViT baseline models for glaucoma detection on a combined dataset of six publicly available fundus images. They also proposed an ensemble of the best ViT models. Swin~\cite{Liu2021Swin} achieved the best standalone performance with 92.57\% in sensitivity, 96.94\% in specificity, and 97.90\% in AUC.
Xu et al.~\cite{xu2021automatic} proposed a Transfer Induced Attention Network (TIA-Net) for automatic glaucoma detection. TIA-Net leverages the fundus feature learned from similar ophthalmic data to extract general features.  
The channel-wise attention and maximum mean discrepancy are then adopted to extract the discriminative features that fully characterize the glaucoma-related deep patterns.
As a result, the proposed method achieves a smooth transition between general and specific features, thus enhancing feature transferability.
Song et al.~\cite{Song2021Deep} proposed a Deep Relation Transformer (DRT) method for diagnosing glaucoma based on the combined OCT and VF modalities. The proposed framework includes three successive modules: the global relation module, the guided regional relation module, and the interaction transformer module. 
These modules utilize deep reasoning and transformer mechanisms to explore implicit pairwise relations between OCT and VF information and enhance the representation with complementary information. The proposed DRT approach outperforms existing methods and has the potential to accurately diagnose glaucoma using multimodal data.

\subsection{Generative Adversarial Networks}

Generative Adversarial Networks (GANs) are a type of deep neural network architecture that has the ability to generate new samples from a given probability distribution~\cite{Goodfellow2014GAN}. The GANs consist of two parts: the generator and the discriminator. The generator is trained to produce realistic samples, while the discriminator is trained to differentiate between synthetic data and real data.
Diaz-Pinto et al.~\cite{Diaz2019Retinal} proposed a framework for glaucoma assessment by developing a fundus image synthesizer and a semi-supervised learning method using the Deep Convolutional Generative Adversarial Network (DCGAN)~\cite{radford2015unsupervised} architecture.
The architecture of the image synthesizer and semi-supervised learning method are identical, except for the last output layer of the discriminator.
The DCGAN discriminator is modified in the semi-supervised learning method to function as a 3-class classifier capable of distinguishing between Normal, Glaucoma, and Real/Fake classes. The models were trained on 86,926 cropped retinal images obtained from fourteen publicly available databases. 

Guo et al.~\cite{guo2022dsln} proposed the use of CycleGAN~\cite{Zhu_2017_ICCV} in a teacher-student framework to reduce the appearance differences between labeled source domain and labeled target domain images, aiming to enhance the accuracy of multiracial glaucoma detection.
The proposed framework consists of an inter-image tutor, an intra-image tutor, a student model and a backbone network, which combines the advantages of domain adaptation and semi-supervised learning. 
The inter-image tutor uses CycleGAN for style transfer and transfers the learned knowledge to the student model by minimizing knowledge distillation loss. This helps to overcome the domain shift problem and improve the performance of glaucoma detection.
The intra-image tutor adopts the exponential moving average to leverage the unlabeled target domain and transfers the knowledge to the student model by minimizing prediction consistency loss. The student model not only directly learns knowledge from the labeled target domain images, but also learns the intra-image knowledge and inter-image knowledge transfer by two tutors. Furthermore, the backbone integrates the context features of the local OD region and global fundus image via modified ResNet-50. Comprehensive experimental results on various datasets are shown in Table~\ref{tab:GAN-Arc}.

\begin{table*}[!htbp]
    \caption{Comparison of papers utilizing GAN-based architectures for glaucoma classification. Studies with (*) mark tested their model on an auxiliary dataset}
    \label{tab:GAN-Arc}
    \resizebox{\textwidth}{!}{%
    \begin{tabular}[!htb]{m{1cm}m{1.5cm}m{1.5cm}m{2cm}m{1cm}m{1cm}m{1cm}m{1.1cm}m{1cm}m{1cm}m{0.75cm}}
        \hline
        \multirow{2}{*}{\textbf{Study}} & \multirow{2}{*}{\textbf{Model}} & \multirow{2}{*}{\textbf{Data type}} & \multirow{2}{*}{\textbf{Dataset}} & \multicolumn{5}{c}{\textbf{Classification Performance Measures (\%)}} & \multirow{2}{*}{\textbf{Code}} \\ \cline{5-9}
         &  & &  & \textbf{ACC} & \textbf{SEN} & \textbf{SPE} & \textbf{F1-score} & \textbf{AUC} &  \\ \hline
        
        \cite{Diaz2019Retinal} & DCGAN & Fundus & 14 publicly available datasets$\color{blue}^{\ddagger}$
        & - & 82.90 & 79.86 & 84.29 & 90.17 & \href{https://figshare.com/s/6e4cbba780b81a59964c}{Code} \\ 
        
        \multirow{8}{2cm}{\cite{guo2022dsln}*} & \multirow{8}{1.5cm}{CycleGAN-based} & \multirow{8}{2cm}{Fundus} & LAG & 98.14 & 98.62 & 98.17 & - & 96.41 & \multirow{8}{2cm}{-} \\
        & & & REFUGE & 98.74 & 96.85 & 96.57 & - & 97.06 & \\
        & & & ORIGA & 97.64 & 97.13 & 97.62 & - & 96.93 & \\
        & & & DRISHTI-GS & 98.45 & 97.66 & 96.84 & - & 97.04 & \\
        & & & ACRIMA & 96.52 & 97.62 & 96.72 & - & 97.26 & \\
        & & & RIM-ONE-r1 & 97.48 & 98.31 & 97.14 & - & 97.60 & \\
        & & & RIM-ONE-r2 & 97.46 & 98.23 & 96.57 & - & 96.40 & \\
        & & & RIM-ONE-r3 & 96.46 & 97.18 & 96.82 & - & 96.56 & \\
        \hline
        
    \multicolumn{11}{m{15.75cm}}{$\color{blue}^{\ddagger}$ ORIGA-light, DRISHTI-GS1, RIM-ONE, sjchoi86-HRF, HRF, DRIVE~\cite{Staal2004Ridge}, MESSIDOR~\cite{WANG201982}, DR KAGGLE~\cite{drkaggle}, STARE~\cite{Hoover2003Locating}, e-ophtha~\cite{DECENCIERE2013196}, ONHSD~\cite{Lowell2004Optic}, CHASEDB1~\cite{Christopher2011Retinal}, DRIONS-DB, SASTRA~\cite{Narasimhan2012Glaucoma}, ACRIMA} \\ 
    
    \end{tabular}%
    }
\end{table*}

\subsection{Geometric Deep Learning Networks}

Geometric deep learning is a machine learning approach that focuses on developing algorithms and network architectures capable of analyzing non-Euclidean structured data, such as graphs, manifolds, and point clouds, by integrating geometric principles and deep learning techniques~\cite{Bronstein2017Geometric}. Thi{\'e}ry et al.~\cite{Thiery2023Medical} proposed using geometric deep learning to diagnose glaucoma from a single OCT scan of the ONH and compared its performance to that of 3D CNN and RNFL thickness. Using a deep learning model, they first segmented seven major neural and connective tissues from OCT images. After that, each ONH was represented as a 3D point cloud with approximately 1,000 points. Geometric deep learning (PointNet~\cite{Charles2017PointNet}) was then used to diagnose glaucoma from a single 3D point cloud. 
The proposed geometric deep learning model achieved an AUC of 95.00\% on a private dataset consisting of 873 glaucomatous and 3,897 non-glaucomatous OCT scans, outperforming that obtained with a 3D CNN (AUC of 87.00\% on raw OCT images and 91.00\% on segmented OCT images) and that obtained from RNFL thickness alone (AUC = 80.00\%).
In another study, Braeu et al.~\cite{Braeu2023Geometric} proposed to compare the performance of PointNet with dynamic graph convolutional neural network (DGCNN)~\cite{Wang2019Dynamic} for diagnosing glaucoma. Following the same procedure as their previous paper~\cite{Thiery2023Medical}, each ONH was represented as a 3D point cloud and used to diagnose glaucoma. They demonstrated that both the DGCNN and PointNet could accurately classify 2,259 glaucomatous from 2,247 Non-glaucomatous OCT scans based on 3D ONH point clouds with an AUC of 97.00\% and 95.00\%, respectively.
Furthermore, they identified critical 3D structural features of the ONH for glaucoma diagnosis, which formed an hourglass pattern mostly located within the NRR in the inferior and superior quadrants of the ONH.

\subsection{Hybrid Networks}

Hybrid networks can help in glaucoma diagnosis by combining the strengths of different types of neural networks. For example, Chai et al.~\cite{CHAI2018147} designed a multi-branch neural network (MB-NN) model to exploit domain knowledge and extract hidden features from retinal fundus images.
The entire fundus image and the OD region image are the inputs to the first and second branches of the MB-NN model, respectively. The OD region was extracted using Faster-RCNN~\cite{NIPS2015_14bfa6bb} trained on a separate dataset. Similarly, they integrated domain knowledge into a one-dimensional feature vector and fed it into the third branch of the proposed model.
Fu et al.~\cite{Fu2018Disc} also proposed a disc-aware ensemble network (DENet) for automatic glaucoma screening that integrates the deep hierarchical context of the global fundus image and the local OD region. The network consists of four deep streams, including a global image stream, a segmentation-guided network, a local Disc region stream, and a Disc polar transformation stream. 
The segmentation-guided network is based on U-shape convolutional network to detect the OD region and guide glaucoma screening on the whole fundus image. The architecture of the other streams is based on the ResNet-50 model. These streams provide complementary information, and their output probabilities are fused to obtain the final classification result. 
Yu et al.~\cite{Yu2020Difficulty} proposed using raw multi-rater gradings to improve the performance of deep learning models for glaucoma classification. Instead of predicting labels from individual raters, the authors proposed a multi-branch structure that generates three predictions with different sensitivity settings for the input images: one with the best sensitivity, one with the best specificity, and one with a balanced fused result.
A consensus loss is introduced to encourage consistent results from the sensitivity and specificity branches for consensus labels and opposite results for disagreement labels.
Meanwhile, the consistency or inconsistency between the prediction results of the two branches is used to determine the difficulty level of an image, which is further used to guide the balanced fusion branch to focus more on hard cases.
Garc{\'\i}a et al.~\cite{GARCIA2021105855} proposed combining CNN and LSTM networks for glaucoma diagnosis from raw SD-OCT volumes. The proposed model consists of a slide-level feature extractor and a volume-based predictive model.
The feature extractor utilized residual and attention convolutional modules combined with fine-tuning techniques. Also, to incorporate spatial information along with the three-dimensional data, the model used the LSTM networks with a sequential-weighting module. This module helps to optimize the LSTM outputs, resulting in a more stable and efficient model learning process.

\begin{table*}[!htbp]
    \caption{Comparison of papers utilizing hybrid architectures for glaucoma classification. Studies with (*) mark tested their model on an auxiliary dataset.
    N: Normal, G: Glaucoma, SINDI: Singapore Indian Eye Study.}
    \label{tab:Hybrid-Arc}
    \resizebox{\textwidth}{!}{%
    \begin{tabular}[!htb]{m{1cm}m{1.5cm}m{1.5cm}m{2cm}m{1cm}m{1cm}m{1cm}m{1.1cm}m{1cm}m{0.75cm}}
        \hline
        \multirow{2}{*}{\textbf{Study}} & \multirow{2}{*}{\textbf{Model}} & \multirow{2}{*}{\textbf{Data type}} & \multirow{2}{*}{\textbf{Dataset}} & \multicolumn{5}{c}{\textbf{Classification Performance Measures (\%)}} & \multirow{2}{*}{\textbf{Code}} \\ \cline{5-9}
         &  & &  & \textbf{ACC} & \textbf{SEN} & \textbf{SPE} & \textbf{F1-score} & \textbf{AUC} &  \\ \hline
        
        \multirow{2}{2cm}{\cite{Fu2018Disc}*} & \multirow{2}{2cm}{DENet} & \multirow{2}{2cm}{Fundus} & ORIGA,
        SCES (46 G, 1,630 N) & 84.29 & 84.78 & 83.80 & - & 91.83 & \multirow{2}{2cm}{\href{https://github.com/HzFu/DENet_GlaucomaScreen}{Code}}
        \\ 
        & & & SINDI (113 G, 5,670 N) & 74.95 & 78.76 & 71.15 & - & 83.80 & \\ 
        
        \cite{CHAI2018147} & MB-NN & Fundus & Private (1,023 G, 1,531 N) & 91.51 & 92.33 & 90.90 & - & - & - \\
        
        \multirow{3}{2cm}{\cite{Yu2020Difficulty}*} & \multirow{3}{1.5cm}{Multi-rater deep model} & \multirow{3}{2cm}{Fundus} & Private (2,952 G, 3,366 N)
        & 92.54 & 90.96 & 93.92 & 91.89 & 97.94 & - \\ 
        & & & DRISHTI-GS1 & 86.14 & 91.43 & 74.19 & 90.14 & 89.63 & - \\ 
        & & & REFUGE & 98.00 & 82.50 & 99.72 & - & 96.83 & - \\ 

        \cite{GARCIA2021105855} & CNN + LSTM & OCT & Private (144 G, 176 N) & 81.25 & 75.86 & 85.71 & 78.57 & 80.79 & - \\
        
        \hline
    \end{tabular}%
    }
\end{table*}

\section{Glaucoma Prediction}
\label{sec:Prediction}
Glaucoma is a degenerative disease that often exhibits minimal symptoms during its initial stages~\cite{Li2022progression}. Therefore, predicting the risk of developing glaucoma is crucial for early detection and timely intervention to prevent vision loss.
Recent studies have demonstrated the potential of deep learning models in predicting the probability of glaucoma development through the analysis of diverse demographic, clinical, and imaging data~\cite{Li2022progression, THAKUR2020262, Dixit2021Assessing}.
Li et al.~\cite{Li2022progression}, for example, proposed to use a CNN-based network to predict and stratify the risk of glaucoma onset and progression based on fundus images of 17,497 eyes in 9,346 patients. The model demonstrated the ability to predict patients who may develop glaucoma within a five-year period with an AUC of 90.00\%. Thakur et al.~\cite{THAKUR2020262} also used deep learning models to predict glaucoma development from fundus images in a prospective longitudinal study several years before disease onset. The study reported an AUC of 77.00\% for predicting the development of glaucoma 4 to 7 years before the onset of the disease, 88.00\% for predicting the development of glaucoma approximately 1 to 3 years before the onset, and 95.00\% for detecting glaucoma after the onset.

Li et al.~\cite{Li2020DeepGF} proposed a deep learning model for glaucoma prediction (DeepGF) based on sequential fundus images. They first established a database of sequential fundus images for glaucoma prediction (SIGF), which included an average of 9 images per eye, for a total of 3,671 images. The proposed DeepGF consists of an attention-polar convolution neural network and a variable time interval long short-term memory (VTI-LSTM) network to learn the spatio-temporal transition at different time intervals across sequential medical images of a person. In addition, a novel active convergence training strategy is proposed to solve the glaucoma forecast's imbalanced sample distribution problem. The proposed method demonstrated an accuracy of 80.7\%, a sensitivity of 85.70\%, a specificity of 80.60\%, and an AUC of 87.00\%.
Dixit et al.~\cite{Dixit2021Assessing} proposed using a convolutional LSTM neural network to detect glaucoma progression from a longitudinal dataset of merged VF and clinical data. The dataset comprises 11,242 eyes with at least four VF results and corresponding baseline clinical data, including CDR, central corneal thickness, and IOP for each sample. The proposed method achieved an AUC of 93.93\% for glaucoma progression prediction.
\cite{Thakur2020Convex} developed a deep archetypal framework to predict glaucoma approximately four years prior to disease onset. The framework utilizes simplex projections to obtain unsupervised convex representations of the visual fields, which proved clinically meaningful and more discriminative than raw VFs or other classical VF analysis approaches. The proposed model achieved an AUC of 71.00\% on a dataset of 7,248 VF tests collected only at the baseline.

\section{Challenges and Future Directions} \label{sec:Challenges}
The adoption of deep learning for glaucoma diagnosis faces several key challenges that need to be addressed through continued research and cross-disciplinary collaboration. We structure these challenges and promising future directions as follows.

\subsection{Data Challenges}
Developing effective deep learning models for glaucoma diagnosis relies on the availability of comprehensive training datasets. However, assembling high-quality annotated datasets presents several key challenges.

One major challenge is the limited size of publicly available glaucoma datasets compared to common computer vision benchmarks like ImageNet. The prohibitive expertise and cost required for reliable manual annotations further constrains dataset development. Data diversity is another issue, with many datasets lacking varied representations across different demographics, ethnicities, and imaging equipment~\cite{DeFauw2018Clinically,Orlando2020REFUGE}. These limitations have caused existing public datasets to remain relatively small in scale, which can hinder model performance and restrict real-world applicability. However, alternative training methods such as transfer learning~\cite{xu2021automatic, latif2022odgnet}, zero/few-shot learning~\cite{garcia2021circumpapillary, wang2020generalizing}, and knowledge distillation~\cite{chelaramani2021multi, ZHAO2022102295, GUO2022106910} offer potential solutions to mitigate the impact of limited dataset size. Additionally, active learning techniques can optimize the annotation process by selectively identifying the most useful and ambiguous samples for labeling. Incorporating uncertainty estimation into the sample selection process~\cite{adhane2022use} further focuses active learning on improving model performance on underrepresented classes.

Imbalanced class distribution poses another persistent problem, with normal cases dominating many current datasets. This can skew model performance toward low sensitivity and high false negative rates~\cite{kaur2019systematic}. Techniques like data augmentation~\cite{shorten2019survey, ASAOKA2019224}, weighted sampling~\cite{cui2019class}, and generation of synthetic minority over-samples~\cite{li2023subspace} may help mitigate imbalance. However, augmentation requires judicious implementation with clinician input to prevent medical inaccuracies and ensure fidelity~\cite{finlayson2019adversarial}.

Healthcare data also carries ethical and legal obligations around privacy that researchers must proactively address~\cite{kondylakis2023data}. Laws and cultural norms around medical data sharing vary geographically, necessitating localized considerations. Policymakers have a role in developing balanced frameworks that promote research and innovation while protecting patient privacy. Techniques like federated learning~\cite{zhang2022splitavg} can also enable collaborative model development without raw data sharing. However, the onus remains on researchers to be privacy stewards and ensure compliance with applicable national and regional privacy laws in their research studies~\cite{vollmer2020machine}.

\subsection{Model Development Challenges}
Deep learning models need to detect subtle signs of early-stage glaucoma, which is challenging even for experienced clinicians~\cite{asaoka2019using}. Heterogeneity in early disease appearance further complicates this task. Researchers are exploring various model architecture techniques to improve performance for early detection. For instance, attention mechanisms have shown promise in focusing models on salient retinal regions, enhancing informative feature extraction, and representation learning~\cite{Li2020Large}.

Integrating multi-modal data and multi-label learning may further improve performance by leveraging interrelated tasks like segmentation and diagnosis~\cite{Song2021Deep}. However, this poses challenges such as optimal data fusion, balancing diverse tasks, and continual learning as new data emerges. To address these issues, researchers are exploring specialized multi-modal architectures, adaptive optimization strategies, dynamic network expansion, and meta-learning~\cite{gao2023discriminative}.

Despite the potential benefits, the opaque nature of deep learning models remains a challenge. The importance of employing or proposing explainable AI (XAI) techniques to explain predictions and feature importance from deep learning models using different image modalities like OCT, Fundus, and VF has been recognized by researchers~\cite{shibata2018development,thompson2020review,hemelings2021deep}. These studies employed post-hoc explanation techniques like SHAP values~\cite{lundberg2017unified}, locally interpretable model-agnostic explanations~\cite{ribeiro2016should}, integrated gradients~\cite{sundararajan2017axiomatic}, occlusion sensitivity~\cite{zeiler2014visualizing}, saliency maps~\cite{vasu2020iterative}, and contrastive explanations~\cite{dhurandhar2018explanations} to identify influential segments, patterns, and abnormalities in fundus images that lead the models to predict glaucoma. They found that XAI methods can highlight relevant regions like the optic disc, retinal nerve fiber layer defects, and areas of hemorrhaging that most inform the models’ predictions~\cite{ting2019deep,mehta2021automated}. However, further development of XAI techniques tailored for glaucoma is still necessary, as most methods are model-agnostic and not optimized for glaucomatous features. Most work has focused on post-hoc methods rather than real-time explainability during model development and evaluation. To enhance understanding and trust in AI-assisted glaucoma screening, the field needs to create glaucoma-specific XAI solutions that provide human-interpretable explanations in real time. Another pragmatic approach involves clinician-researcher collaboration to assess XAI techniques and identify medically relevant insights~\cite{Liao2020Clinical}.





\subsection{Clinical Translation Challenges}
For clinical adoption, glaucoma diagnosis models must demonstrate consistent performance across diverse populations and imaging equipment. However, demographic and acquisition differences can affect model generalizability. Wider collaborations between academia, healthcare providers, and industry could facilitate large-scale external validation across multiple centers and populations to identify failure modes. The models should integrate seamlessly into ophthalmic workflows with efficient computations on edge devices~\cite{Rahmani2018Exploiting}. Physician trust is integral for uptake, requiring initiatives to improve model explainability and transparency. Frameworks for uncertainty estimation could also indicate situations where clinician oversight is necessary.

Careful design of human-AI interaction mechanisms will be vital, allowing physicians to accept, reject or modify model recommendations. Regulatory agencies play a crucial role in establishing standards for rigorous clinical validation of deep learning systems before approval~\cite{kelly2019key}. Beyond accuracy, aspects like usability, interoperability, cybersecurity and patient privacy must be addressed to ensure safety and effectiveness~\cite{matheny2020artificial}. Overall, a patient-centered approach with clinician partnership in the model development lifecycle will be key for clinical translation.

\subsection{Future Outlook}
Advancements in glaucoma diagnosis will require synergistic progress across multiple disciplines. On the data front, innovations in sensor and imaging technologies can enable the acquisition of informative multi-modal datasets. In parallel, advancements in deep learning architectures, optimization algorithms, and computing hardware will allow more sophisticated analysis of these rich datasets.

Several promising research directions could accelerate translating of these innovations into clinical practice. Meta-learning approaches may enable rapid model adaptation from limited annotated examples, mitigating data constraints. Adversarial techniques can improve model robustness to input perturbations. Reinforcement learning offers the potential for optimizing glaucoma management policies. Recurrent neural networks could integrate longitudinal patient data for enhanced monitoring.

Ultimately, sustained collaborative efforts spanning medicine, engineering and computer science will be key to realizing artificial intelligence's potential for enhancing glaucoma outcomes. Initiatives to promote open datasets, model repositories and evaluation benchmarks will facilitate collective progress. Interdisciplinary teams should lead technological development, ensuring clinical applicability and integration into practice. With concerted efforts, AI-enabled glaucoma care could soon transition from promise to reality.

\section{Conclusion} \label{sec:Conclusion}
In this paper, we have provided a comprehensive overview of the state-of-the-art research applying deep learning and computer vision techniques for glaucoma diagnosis. Through a systematic literature review methodology, we synthesized the existing work across diverse architectural categories, including CNNs, autoencoders, attention networks, GANs, and geometric deep learning models. 

The review highlighted promising capabilities demonstrated by these techniques in analyzing fundus, OCT, and visual field data. Tasks like classification, segmentation, and prediction of glaucoma have shown strong results across a wide range of experiments. We also discussed different feature extraction approaches, covering structural, statistical, and hybrid techniques for identifying informative glaucoma biomarkers from retinal imaging data. However, key challenges remain around limited dataset size and diversity, class imbalance, optimizing models for early disease detection, integrating multi-modal data, and translating solutions to clinical practice.

Ongoing efforts are beginning to address these gaps through transfer learning approaches, data augmentation techniques, attention mechanisms, multi-task learning, model explainability, and physician collaboration. Nonetheless, realizing the full potential of AI in transforming glaucoma care will require sustained cross-disciplinary teamwork. 

From our perspective, open datasets, model repositories, and evaluation benchmarks will be critical to accelerate collective progress. Ultimately, an integrated approach spanning medicine, engineering and computer science will be essential for developing and validating solutions ready for real-world clinical deployment. This review highlights the tremendous opportunities for us at the intersection of ophthalmology and artificial intelligence. Overall, we aimed to provide a comprehensive overview and analysis of the state-of-the-art in this exciting and high-impact emerging field.

\section*{Acknowledgment}
\noindent \textbf{Funding:} This work was supported in part by the European Research Council (ERC) through the Horizon 2020 research and innovation program under grant agreement number 101002711. Mohammad Mahdi~Dehshibi received partial funding from this source.

\noindent \textbf{Author Contributions:} All authors contributed to the conception of the idea, defined the scope of the survey, and developed the survey methodology. Mona~Ashtari-Majlan conducted the literature review, identified research gaps, analyzed survey data, and wrote the paper. All authors participated in the discussion of results, provided feedback on the manuscript, and assisted in writing and editing.

\noindent \textbf{Competing Interests:} The authors declare no competing interests.

\noindent \textbf{Data and Materials Availability:} All data presented in this study is available within the main text.

\bibliographystyle{ieeetr}
\bibliography{references,mohammad_ref}

\vspace*{-2\baselineskip}
\begin{IEEEbiography}[{\includegraphics[width=1in,height=1.25in,clip,keepaspectratio]{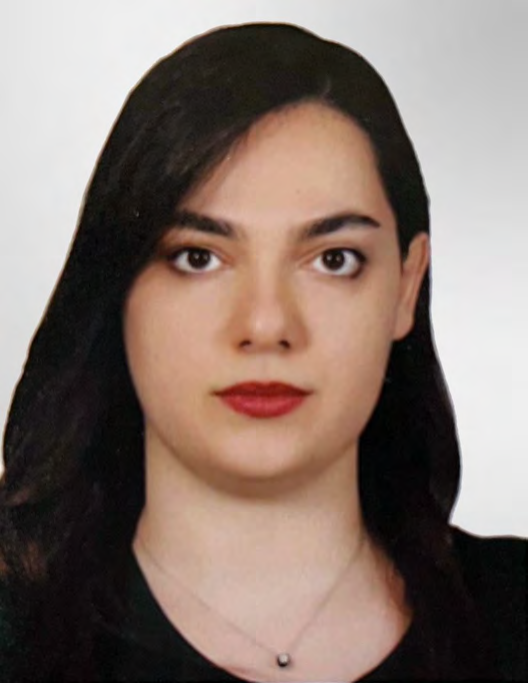}}]{Mona Ashtari-Majlan}
received her Master’s degree in Health Systems Engineering from Amirkabir University of Technology, Tehran, in 2021.
She is currently a Ph.D. candidate in the Doctoral program in Network and Information Technologies at Universitat Oberta de Catalunya, Spain. Her research interests include Biomedical Image Processing, Computer Vision, and Deep Learning.

\end{IEEEbiography}
\vspace*{-2\baselineskip}
\begin{IEEEbiography}[{\includegraphics[width=1in,height=1.25in,clip,keepaspectratio]{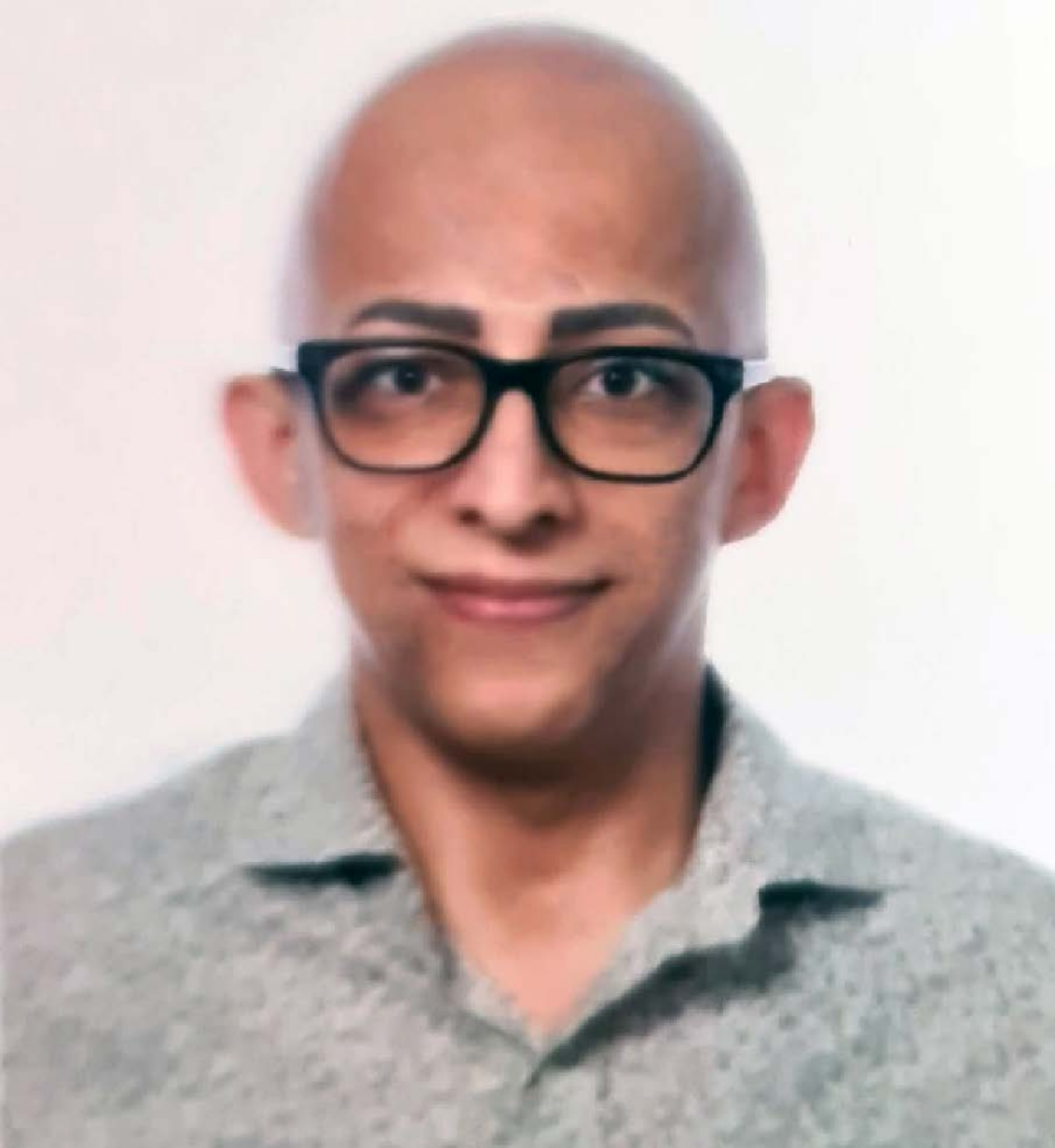}}]{Mohammad Mahdi Dehshibi}
received his Ph.D. in Computer Science in 2017. He is currently a research scientist at Universidad Carlos III de Madrid, Spain. He is also an adjunct researcher at Universitat Oberta de Catalunya (Spain) and the Unconventional Computing Lab. at UWE (Bristol, UK). He has contributed to more than 70 papers published in peer-reviewed journals and conference proceedings. He also serves as an associate editor of the \emph{International Journal of Parallel, Emergent, and Distributed Systems}. His research interests include Deep Learning, Medical Image Processing, Affective Computing, and Unconventional Computing.
\end{IEEEbiography}
\vspace*{-2\baselineskip}
\begin{IEEEbiography}[{\includegraphics[width=1in,height=1.25in,clip,keepaspectratio]{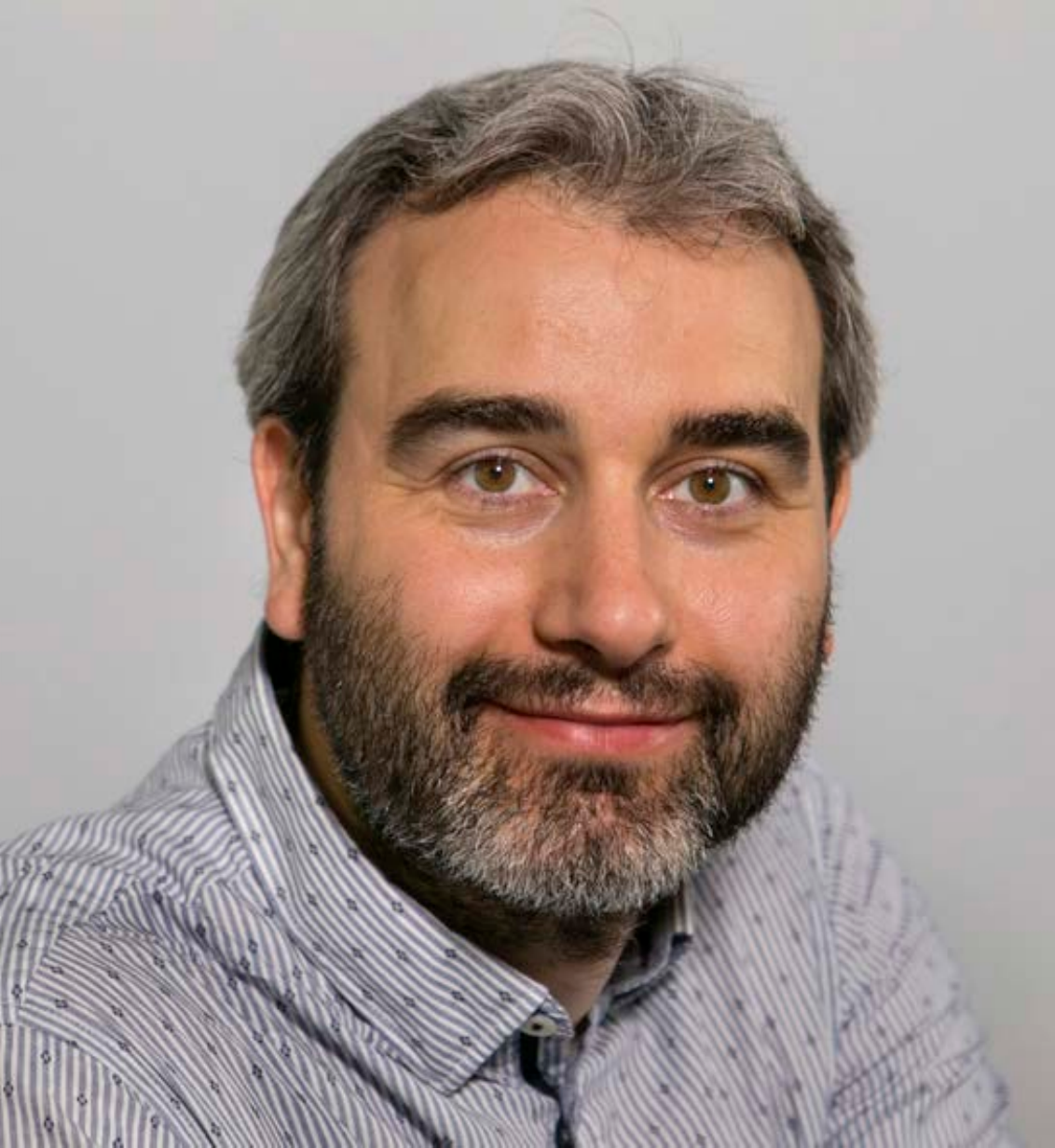}}]{David Masip} (Senior Member, IEEE) received his Ph.D. degree in Computer Vision in 2005 (Universitat Autonoma de Barcelona, Spain). He was awarded for the best thesis in Computer Science. He is a Full Professor at the Computer Science Multimedia and Telecommunications Department at Universitat Oberta de Catalunya, Spain, and the Director of the Doctoral School since 2015. He has published more than 70 scientific papers in relevant journals and conferences. His research interests include Oculomics, Retina Image Analysis and Affective Computing.
\end{IEEEbiography}
\vfill

\end{document}